\documentclass[aps,prb,superscriptaddress,twocolumn,showpacs]{revtex4}

\usepackage{graphicx}
\usepackage{dcolumn}
\usepackage{bm}

\begin{document}

\title{Magnetization, specific heat, and thermal conductivity of hexagonal ErMnO$_3$ single crystals}

\author{J. D. Song}
\affiliation{Hefei National Laboratory for Physical Sciences at Microscale, University of Science and Technology of China, Hefei, Anhui 230026, People's Republic of China}

\author{C. Fan}
\affiliation{Hefei National Laboratory for Physical Sciences at Microscale, University of Science and Technology of China, Hefei, Anhui 230026, People's Republic of China}

\author{Z. Y. Zhao}
\affiliation{Hefei National Laboratory for Physical Sciences at Microscale, University of Science and Technology of China, Hefei, Anhui 230026, People's Republic of China}
\affiliation{Fujian Institute of Research on the Structure of Matter, Chinese Academy of Sciences, Fuzhou, Fujian 350002, People's Republic of China}

\author{F. B. Zhang}
\affiliation{Hefei National Laboratory for Physical Sciences at Microscale, University of Science and Technology of China, Hefei, Anhui 230026, People's Republic of China}

\author{J. Y. Zhao}
\affiliation{Hefei National Laboratory for Physical Sciences at Microscale, University of Science and Technology of China, Hefei, Anhui 230026, People's Republic of China}

\author{X. G. Liu}
\affiliation{Hefei National Laboratory for Physical Sciences at Microscale, University of Science and Technology of China, Hefei, Anhui 230026, People's Republic of China}

\author{X. Zhao}
\affiliation{School of Physical Sciences, University of Science and Technology of China, Hefei, Anhui 230026, People's Republic of China}

\author{Y. J. Liu}
\affiliation{School of physics, Huazhong University of Science and Technology, Wuhan, Hubei 430074, People's Republic of China}
\affiliation{Wuhan National High Magnetic Field Center, Huazhong University of Science and Technology, Wuhan, Hubei 430074, People's Republic of China}

\author{J. F. Wang}
\affiliation{School of physics, Huazhong University of Science and Technology, Wuhan, Hubei 430074, People's Republic of China}
\affiliation{Wuhan National High Magnetic Field Center, Huazhong University of Science and Technology, Wuhan, Hubei 430074, People's Republic of China}

\author{X. F. Sun}
\email{xfsun@ustc.edu.cn}

\affiliation{Hefei National Laboratory for Physical Sciences at Microscale, University of Science and Technology of China, Hefei, Anhui 230026, People's Republic of China}

\affiliation{Key Laboratory of Strongly-Coupled Quantum Matter Physics, Chinese Academy of Sciences, Hefei, Anhui 230026, People's Republic of China}

\affiliation{Collaborative Innovation Center of Advanced Microstructures, Nanjing, Jiangsu 210093, People's Republic of China}

\date{\today}

\begin{abstract}

We report a study of magnetism and magnetic transitions of hexagonal ErMnO$_3$ single crystals by magnetization, specific heat and heat transport measurements. Magnetization data show that the $c$-axis magnetic field induces three magnetic transitions at 0.8, 12 and 28 T. The specific heat shows a peak at 2.2 K, which is due to a magnetic transition of Er$^{3+}$ moments. For low-$T$ thermal conductivity ($\kappa$), a clear dip-like feature appears in $\kappa(H)$ isotherm at 1--1.25 T for $H \parallel ab$; while in the case of $H \parallel c$, a step-like increase is observed at 0.5--0.8 T. The transition fields in $\kappa(H)$ are in good agreement with those obtained from magnetization, and the anomaly of $\kappa$ can be understood by a spin-phonon scattering scenario. The natures of magnetic structures and corresponding field-induced transitions at low temperatures are discussed.

\end{abstract}

\pacs{66.70.-f, 75.47.-m, 75.50.-y, 75.85.+t}

\maketitle

\section{INTRODUCTION}

Multiferroicity has attracted many research interests due to its potential applications in magnetoelectronics, spintronics and magnetic memory technology.\cite{Tokura06, materialstoday07, AM10, Physics12, JPCM15} Hexagonal($h$-) manganites $R$MnO$_3$ ($R =$ Y and rare-earth elements) have been found to be typical examples of type-I multiferroics, of which ferroelectricity and magnetism originate from different physical mechanisms, usually leading to a rather weak magnetoelectric coupling.\cite{Khomskii09, WangKF09} This is the main limitation for the practical usage of type-I
multiferroics. Whereas, recent experiments have confirmed $h$-$R$MnO$_3$ being the special one of type-I multiferroics, which exhibits significant magnetoelectric coupling between the $c$-axis polarization and the $ab$-plane staggered antiferromagnetic (AF) magnetization.\cite{Nature02, Nature04, NM04, Nature08} So far, a plenty of works, both theoretical and experimental, have been systematically carried out to discover the hidden multiferroic physics in $h$-$R$MnO$_3$.\cite{Sharma04, Petit07, Poirier07, Nature08, Fabreges09, Jang10, Wang10, Wang12, Choi13, Basistyy14, Toulouse14, Das14, Artyukhin14, Cano14, SunQC14, WangX14, Gupta15, Lilienblum15, YeM15, Paul15} In particular, some special experimental techniques such as high-resolution neutron diffraction, synchrotron X-ray diffraction,\cite{Nature08, Fabreges09} inelastic neutron scattering,\cite{Petit07,Gupta15} femtosecond pump-probe differential reflectance spectroscopy,\cite{Jang10} optical transmission spectroscopy,\cite{Basistyy14} Raman and terahertz spectroscopies,\cite{Toulouse14} ultrasound measurement,\cite{Poirier07} thermal conductivity measurement,\cite{Sharma04, Wang10, Wang12, Choi13} etc., all revealed a strong spin-lattice coupling in $h$-$R$MnO$_3$, which may play the key role in generating the strong magnetoelectric coupling. Meanwhile, the magnetic properties of $h$-$R$MnO$_3$ are rather complicated because of their special crystal structure. 

These materials are crystallized in the hexagonal lattice with the space group $P6_3cm$ at room temperature.\cite{Acta1955, Alonso2000} The Mn$^{3+}$ ions locate at ``6$c$" positions with symmetry $m$ and the $R^{3+}$ ions locate at ``2$a$" and ``4$b$" positions with symmetry 3$m$ and 3, respectively. Besides, the Mn$^{3+}$ ions are arranged into triangular-lattice layer in the $ab$ plane with AF nearest-neighbor interaction. Therefore, the geometrical frustration of magnetic moments is introduced in $h$-$R$MnO$_3$. Once the $R$ sites are occupied by rare-earth magnetic ions like Ho$^{3+}$, Er$^{3+}$, Yb$^{3+}$ etc., the multifold ordering and the 3$d$-4$f$ interaction between the $R^{3+}$ and Mn$^{3+}$ sublattices accompanied with the inherent geometrical frustration can give rise to complicated magnetic phase diagrams and magnetic transitions.\cite{Fiebig03, Yen05, Meier12, Choi13} It is known that the magnetic structures of $R^{3+}$ or Mn$^{3+}$ in the ground state have six possible space groups labeled with $\Gamma_i$ ($i =$ 1--6).\cite{Munoz2000, Fiebig03, Fiebig06, WangYT13, Lorenz13, Sim16} However, there is great difficulty in experimentally distinguishing these spin configurations, resulting in an obstacle in understanding the low-$T$ magnetic structures and the transitions induced by the magnetic field. In particular, it is rather unclear or controversial on the magnetisms and spin configurations of the rare-earth sublattices. Because the Mn$^{3+}$ moments form better magnetic ordering compared to the rare-earth moments, the total magnetic signal of some symmetry sensitive methods like neutron scattering and magnetic Second Harmonic Generation (SHG) method is dominated by the Mn$^{3+}$ moments, leading to much difficulty in determining the magnetic contribution of $R^{3+}$ ions using these direct experimental techniques.\cite{Meier12} But in most cases, the anisotropic measurements like magnetization, thermal expansion, etc., all supported an easy-axis magnetic ordering along the $c$ axis for the rare-earth sublattice.\cite{Yen05, Meier12, Lorenz13, Sugie02, Fiebig02, Yen07, Cruz05, Lorenz05} This kind of anisotropy can be explained by the $3d-4f$ interaction between Mn$^{3+}$ and $R^{3+}$ moments, that is, the $R^{3+}$ moments can form magnetic ordering in the Mn$^{3+}$ molecular field and their orientations are clearly coupled to the Mn$^{3+}$ ones through an energy term of anisotropic nature.\cite{Fabreges08, Fabreges09} Moreover, because the $R^{3+}$(2a) and $R^{3+}$(4b) are located at different Mn$^{3+}$ environments, the magnetic ordering of the $R^{3+}$(4b) can be triggered by the Mn$^{3+}$ ordering at a rather high temperature via a strong 3$d$-4$f$ interaction. In contrast, the coupling between the Mn$^{3+}$ and $R^{3+}$(2a) moments is negligible in comparison with the $R^{3+}$(2a)-$R^{3+}$(2a) interaction, and the magnetic ordering of $R^{3+}$(2a) usually occurs at very low temperatures. Among the set of irreducible representations $\Gamma_1$ to $\Gamma_6$, the four one-dimensional representations $\Gamma_1$ to $\Gamma_4$, as shown in Fig. 1, are always sufficient to describe the magnetic ordering of $R^{3+}$ moments (parallel/antiparalle to $c$ axis), while the two-dimensional representations $\Gamma_5$ and $\Gamma_6$ with the $ab$-plane ordering and the combinations of representations with lower symmetries are not needed.\cite{Meier12}

\begin{figure}
\includegraphics[clip,width=8.5cm]{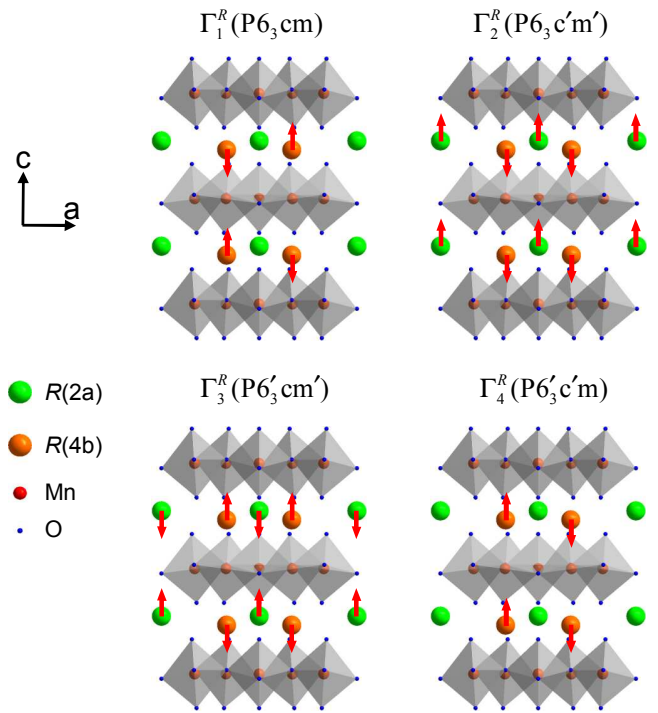}
\caption{(color online) Possible magnetic structures with moments parallel/antiparallel to the $c$ axis for rare-earth sublattice in $h$-$R$MnO$_3$ denoted by four one-dimensional representations $\Gamma_1$ to $\Gamma_4$.}
\end{figure}

Here, we focus on the magnetic properties and magnetic transitions of hexagonal ErMnO$_3$. It is known that ErMnO$_3$ undergoes a ferroelectric transition at $T_c =$ 833 K and keeps the space group $P6_3cm$ below $T_c$. The magnetic ground state of ErMnO$_3$ has been studied by using magnetization, optical second harmonic generation and neutron diffraction measurements.\cite{Sugie02, Fiebig02, Yen07, Meier12, Lorenz13, Chaix14} It was proposed that: the Mn$^{3+}$ moments form an AF order with a $P6'_3c'm$ spin structure at $T_N =$ 79 K; simultaneously, the Er$^{3+}$(4b) moments form the magnetic order with the same $P6'_3c'm$ spin structure at $T_N$; with lowering temperature, the Er$^{3+}$(2a) moments order into a $P6_3c'm'$ spin structure at 10 K with spin alignment in the $c$ axis; the long-range order of Er$^{3+}$(2a) moments at low temperatures can induce a second magnetic transition of Er$^{3+}$(4b) and Mn$^{3+}$ due to the 4$f$-4$f$ and 3$d$-4$f$ magnetic interactions,\cite{Meier12} that is, the Er$^{3+}$(4b) moments can transform from $P6'_3c'm$ to $P6_3c'm'$ accompanied with the synchronous transition of the Mn$^{3+}$ moments from $P6'_3c'm$ to $P6_3c'm'$ at 2 K. Finally, the ground-state magnetic structures of Er$^{3+}$(2a), Er$^{3+}$(4b) and Mn$^{3+}$ sublattices were all proposed to possess the symmetry of $P6_3c'm'$.\cite{Meier12} With applying a magnetic field along the $c$ axis in the ground state, the Er$^{3+}$ moments undergo a ferromagnetic transition from ferrimagnetic arrangement to a full polarization.\cite{Meier12}

In this paper, we report a systematic study of magnetization, specific heat and low-$T$ heat transport of ErMnO$_3$ single crystals at low temperatures and in high magnetic fields. It is found that the magnetization for $H \parallel c$ displays three magnetic transitions at 0.8, 12 and 28 T, respectively. Apparently, the previously proposed ferrimagnetic $P6_3c'm'$ ground state of Er$^{3+}$ sublattices cannot be used to explain these magnetic transitions. Combining the results of magnetization, specific heat, and thermal conductivity, we discuss the ground state and magnetic-field-induced transitions.

\section{EXPERIMENTS}

High-quality ErMnO$_3$ single crystals were grown by the floating-zone technique.\cite{Fan JCG} Chemical compositions of the as-grown crystals were checked by X-ray fluorescence pattern (XRF). The samples used for XRF were cut along the cross section of the single-crystal rod from five different positions with a thickness about 1.5 mm. An averaged molar ratio of Er: Mn in these crystals was confirmed to be 0.99: 1, which is very close to the nominal composition. To confirm the phase purity, some single-crystal pieces were ground into powder to be characterized by X-ray diffraction (XRD). As shown in Fig. 2(a), the powder diffractions show a pure phase of these crystals. Moreover, the phase and single crystallinity of the crystals were also checked by the Laue photograph and X-ray rocking curve measurements. As the representative data, Figs. 2(b) and (c) show the $(00h)$ Bragg peak and the rocking curve of (004) for, respectively. Very narrow width of the rocking curve (FWHM = $0.06^\circ$) demonstrates the good crystallization of the crystals.

\begin{figure}
\includegraphics[clip,width=8.5cm]{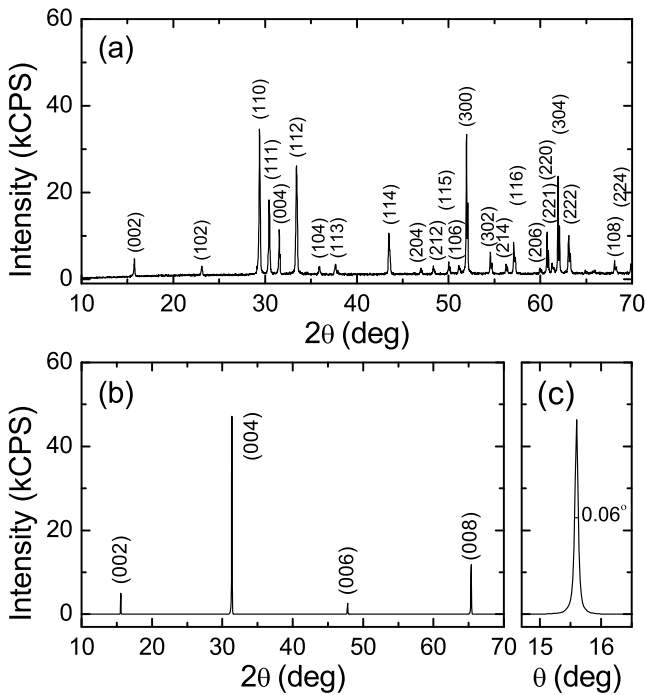}
\caption{(color online) (a) Powder X-ray diffraction of ErMnO$_3$ single crystals. (b,c) X-ray diffraction pattern of $(00h)$ plane and the rocking curve of (004) peak. The full width at the half maximum (FWHM) of the rocking curve is shown in panel (c).}
\end{figure}

The samples used for magnetization, specific heat and thermal conductivity measurement were carefully checked by X-ray Laue photograph and cut precisely along different crystallographic axes. The magnetization was measured in a commercial SQUID-VSM (Quantum Design) and a self-built pulsed magnetic-field platform. The specific heat was measured in a commercial physical property measurement system (PPMS, Quantum Design). The thermal conductivity was measured by using a conventional steady-state technique in a $^3$He refrigerator equipped with a 14 T magnet at the temperature range of 0.3--30 K, and a $^4$He pulse-tube refrigerator for zero-field data above 4 K.\cite{Wang10, Wang12, Zhao_GFO, Zhao_DFO} All these measurements were carried out with increasing temperature of magnetic field after cooling the samples in zero field.

\section{RESULTS AND DISCUSSION}

\subsection{Magnetization}

\begin{figure}
\includegraphics[clip,width=6.5cm]{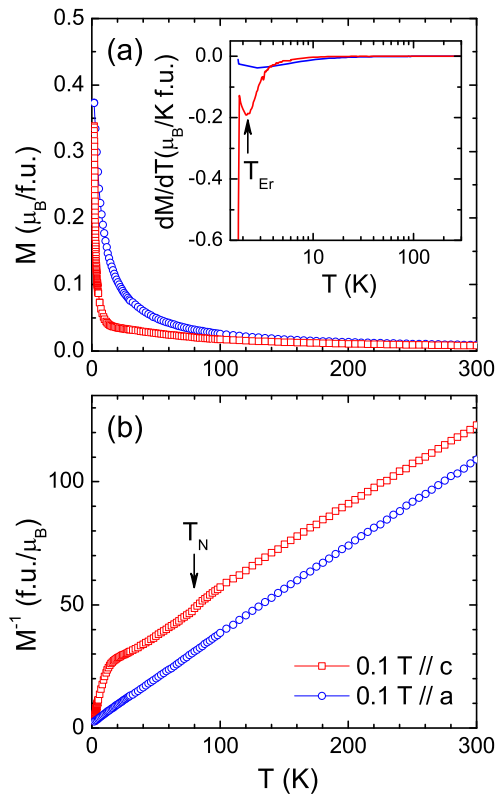}
\caption{(color online) (a,b) Temperature dependence of the magnetization and the inverse magnetization of ErMnO$_3$ for magnetic field (0.1 T) applied along the $c$ and $a$ axes. Inset to panel (a): temperature dependence of $dM/dT$. The minimum at 2.2 K indicates the magnetic-order transition of Er$^{3+}$(2a) moments. In panel (b), the arrow indicates the slope change of $M^{-1}(T)$ at $\sim$ 79 K, which should be corresponding to the N\'eel temperature of Er$^{3+}$(4b) and Mn$^{3+}$ moments. }
\end{figure}

Figure 3 shows the temperature dependence of the magnetization and inverse magnetization along the $c$ and $a$ axes at 0.1 T. As shown in Fig. 3(b), the $M^{-1}(T)$ curve with $H \parallel c$ exhibits a clear change of slope at $T_N =$ 79 K while that with $H \parallel a$ displays a simply paramagnetic behavior, which agree well with previous report.\cite{Lorenz13} It is already known that the Mn$^{3+}$ moments have easy-plane anisotropy in the triangular-lattice $ab$ plane and form an AF long-range order at $T_N$, in which the neighboring moments are rotated by $120^{\circ}$ and can be canceled out.\cite{Kawamura98, Fiebig02, Meier12} Meanwhile, because of a strong 3$d$-4$f$ interaction between the Mn$^{3+}$ and Er$^{3+}$(4b) moments, the ordering of Mn$^{3+}$ sublattice can drive Er$^{3+}$(4b) moments to order into a $P6'_3c'm$ state with spin alignment along the $c$ axis.\cite{Meier12} With lowering temperature, as shown in the inset to Fig. 3(a), the $dM/dT$ curve shows up a minimum at 2.2 K for $H \parallel c$, indicating a magnetic transition of the Er$^{3+}$ moments.

\begin{figure}
\includegraphics[clip,width=8.5cm]{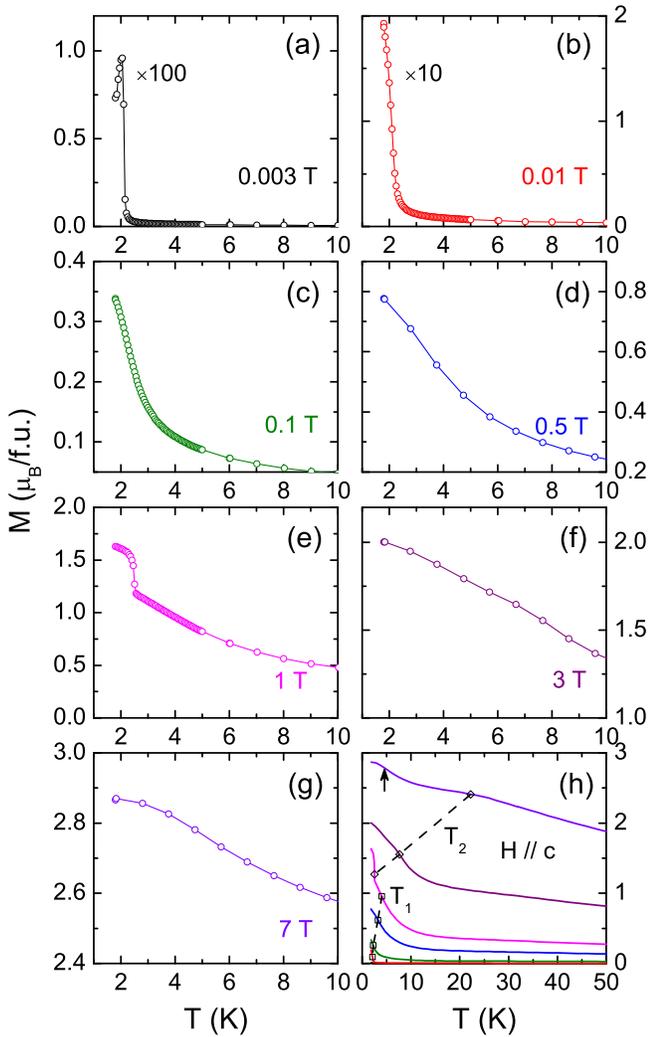}
\caption{(color online) (a--g) Temperature dependence of the magnetization of ErMnO$_3$ in the $c$-axis magnetic fields. (h) A comparison among all $M(T)$ curves. The squares and diamonds indicate two anomalies, respectively, and the dashed lines show the evolution of the anomalies with increasing field, which are in good agreement with previous reports.\cite{Meier12, Lorenz13} Besides, the arrow indicates another anomaly of $M(T)$ in 7 T field. All the anomalies are determined from $dM/dT$.}
\end{figure}

Figure 4 shows the evolution of $M(T)$ with changing fields for $H \parallel c$. Similar data can be found in an earlier literature, which were obtained with both increasing and decreasing temperatures after field cooling.\cite{Meier12} It is found that the $M(T)$ curve in a field of 0.003 T shows a clear peak around 2 K, as shown in Fig. 4(a). It seems that this peak-like feature is due to an AF transition, and the magnetic ground state of ErMnO$_3$ is a kind of AF states. However, by slightly increasing field to 0.01 T, the peak-like feature disappears. It would be strange if an AF state could be destroyed by such a small magnetic field. Figure 4(h) shows the evolution of $M(T)$ curves with changing field. It can be easily found by using the derivative data $dM/dT$ (not shown here) that the $M(T)$ curve for $\mu_0H \leq$ 1 T shows up a minimum, which defines a transition temperature $T_1$. With increasing field, $T_1$ slightly moves to higher temperatures. Moreover, the 1 T curve shows up a step-like increase at $T_2 =$ 2.2 K, which should be associated with a field-induced ferromagnetic transition of the Er$^{3+}$ moments, and the transition temperature $T_2$ shifts quickly to higher temperature with increasing field. These results agree well with previous reports.\cite{Meier12, Lorenz13} In addition, it is notable that in the $M(T)$ for 7 T $\parallel c$ another ferromagnetic-like transition is observed at 4.8 K, as shown in Fig. 4(h). Therefore, the ground-state magnetic structure of ErMnO$_3$ is confirmed to experience at least two magnetic transitions in a $c$-axis field. In this regard, the ferrimagnetic $P6_3c'm'$ spin structure proposed previously can not be the actual magnetic ground state of Er$^{3+}$ moments in ErMnO$_3$.\cite{Meier12}

\begin{figure}
\includegraphics[clip,width=8.5cm]{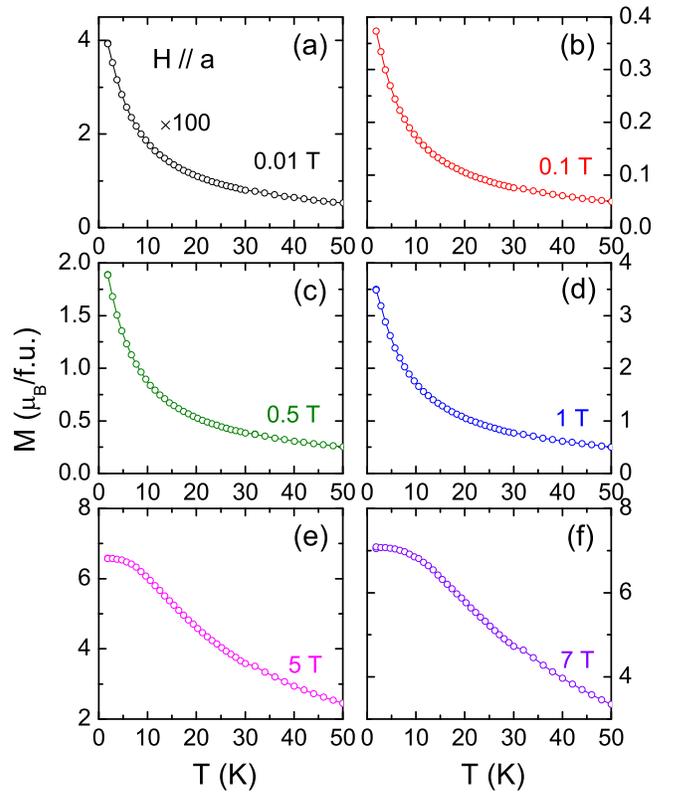}
\caption{(color online) Temperature dependence of the magnetization of ErMnO$_3$ single crystals in the $a$-axis magnetic fields.}
\end{figure}

In the case of $H \parallel a$, as shown in Fig. 5, there is no clear anomaly in $M(T)$ curves for the fields up to 7 T. The magnetization in the $a$-axis field looks like a simply paramagnetic behavior.

\begin{figure}
\includegraphics[clip,width=8.5cm]{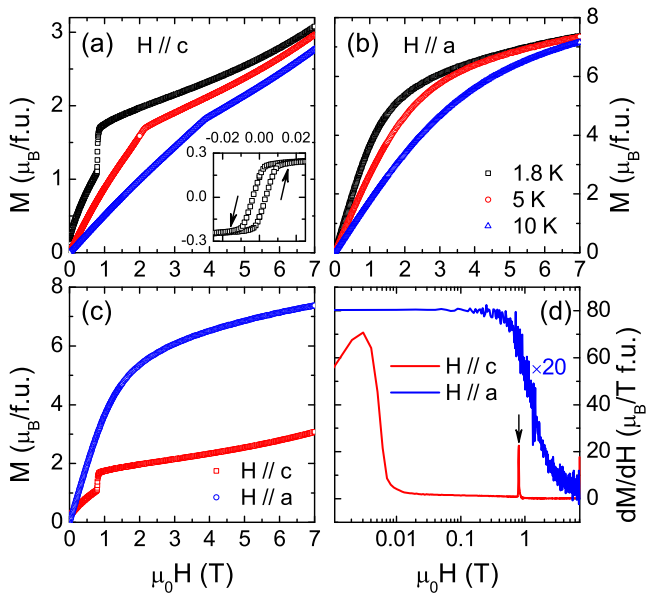}
\caption{(color online) (a,b) Low-temperature magnetization curves $M(H)$ of ErMnO$_3$ in magnetic fields along the $c$ and $a$ axes, respectively. (c) A comparison between the $c$- and $a$-axis magnetization. (d) The differential results of $dM/dH$. Inset to panel (a): low-field data of the magnetization at 1.8 K that exhibit a small hysteresis. The arrows indicates the direction of sweeping field. There is no any hysteresis in all other curves.}
\end{figure}

Figure 6 shows the magnetization curves at low temperatures with magnetic fields along the $a$ or $c$ axis. For $H \parallel c$, the $M(H)$ curve at 1.8 K shows two step-like increases at 0.003 and 0.8 T, which agrees well with the previous reports.\cite{Meier12, Lorenz13} With increasing temperature, the low-field transition disappears and the high-field one moves to higher fields. Here, a hysteresis loop can be seen at the low-field transition with a very small coercive field and remnant magnetization. This feature has a good correspondence to the peak-like feature in the $M(T)$ curve. Note that such a small remnant magnetization cannot be associated with normal ferromagnetism or ferrimagnetism, and is probably caused by a kind of weak ferromagnetism due to spin canting.\cite{Liu, Pato} Increasing temperature to 5 K, this small hysteresis disappears. Similar results have been found by Sugie {\it et al.}\cite{Sugie02} Nevertheless, such small hysteresis loop is incompatible with the previously proposed ferrimagnetic $P6_3c'm'$ ground state of Er$^{3+}$ sublattices.\cite{Meier12} Besides, it is found that the high-field increase also has good correspondence with the transition at $T_2$ in the $M(T)$ curve for $H \parallel c$. In contrast, the $M(H)$ for $H \parallel a$ behaves more simply, which is in good accordance with the results of $M(T)$. However, the slope of $M(H)$ for $H \parallel a$ at 1.8 K shows significant difference before and after 1 T, indicating a possible magnetic transition of Er$^{3+}$ moments. Figure 6(c) shows the comparison between the $c$- and $a$-axis $M(H)$ curves at 1.8 K. Because the Mn$^{3+}$ moments are $120^{\circ}$ ordered in the $ab$ plane, the net moments of total Mn$^{3+}$ ions can be ignored in comparison with that of the Er$^{3+}$ ions. Therefore, the low-$T$ magnetization is mainly contributed by the Er$^{3+}$ moments. It is a puzzle the magnetization for $H \parallel a$ is much larger than that for $H \parallel c$, although the spin-easy axis of rare-earth ions is believed to be along the $c$ axis due to the existence of a $c$-axis molecular field from the Mn$^{3+}$ moments.\cite{Meier12, Chaix14} A possible reason is related to the quadrupolar charge-density distributions of 4$f$ shell of rare-erath ions, as proposed by Skumryev {\it et al.}\cite{Skumryev09}

\begin{figure}
\includegraphics[clip,width=8.5cm]{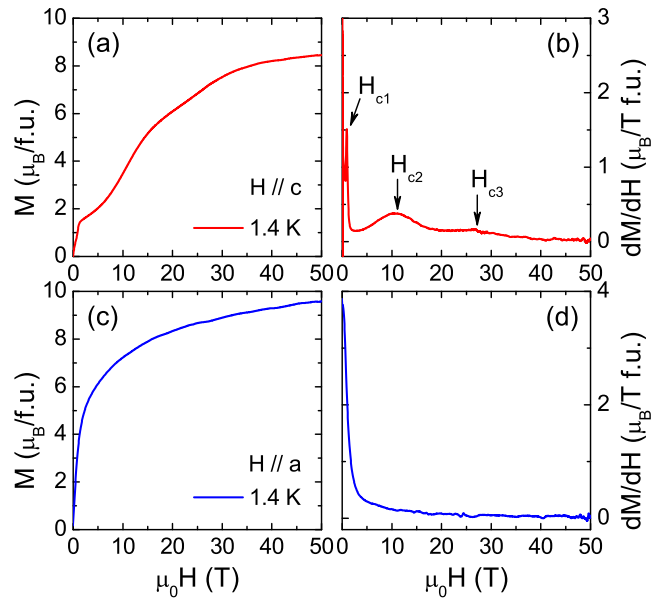}
\caption{(color online) (a,c) Low-temperature magnetization curves $M(H)$ of ErMnO$_3$ obtained in pulsed high magnetic field along the $c$ and $a$ axes. (b,d) The differential results of $dM/dH$ for the $c$- and $a$-axis $M(H)$. The arrows in panel (b) indicate three magnetic transitions at 0.8, 12 and 28 T.}
\end{figure}

The magnetization was also measured in pulsed high magnetic fields up to 50 T. As shown in Fig. 7(a), the $M(H)$ curve with $H \parallel c$ at 1.4 K shows three anomalies, which are clearer in the differential data. The $dM/dH$ curve shown in Fig. 7(d) exhibits three maximums at 0.8, 12 and 28 T (denoted by $H_{c1}$, $H_{c2}$ and $H_{c3}$, respectively). This indicates that the ground-state magnetic structure of ErMnO$_3$ can experience three field-induced transitions for $H \parallel c$. In the case of $H \parallel a$, the $M(H)$ curve at 1.4 K shows a steady increase with increasing field, but the slope of $M(H)$ undergoes a clear change at 1--2 T.

\subsection{Specific heat}

\begin{figure}
\includegraphics[clip,width=6.5cm]{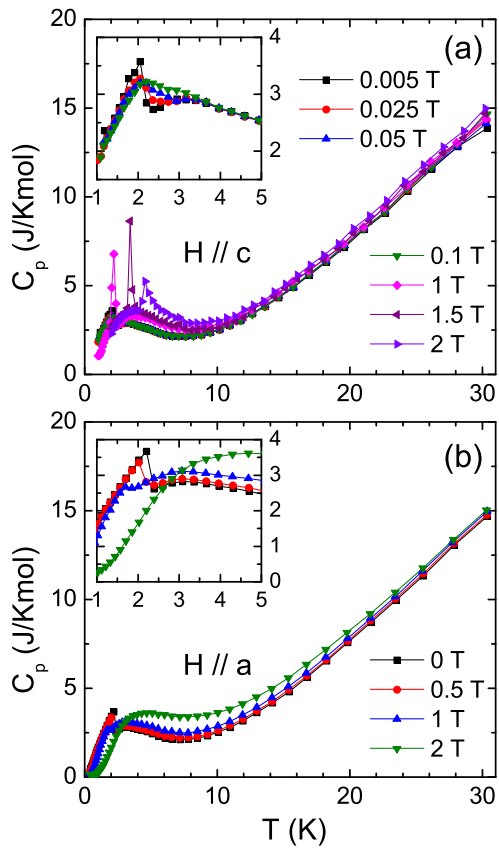}
\caption{(color online) Low-temperature specific heat of ErMnO$_3$ single crystals with magnetic fields along the $c$ or $a$ axis. The insets show the low-temperature data.}
\end{figure}

The low-$T$ specific heat of ErMnO$_3$ single crystals in magnetic fields parallel to the $c$ or $a$ axis are shown in Fig. 8. Here, the zero-field data are consistent with the results in literature.\cite{Lorenz13, Skumryev09} It is found that the zero-field specific heat significantly deviates from the pure-phonon behavior below 10 K. With further lowering temperature, a hump-like feature is observed around 3 K. In addition, a clear peak is observed at 2.2 K, which corresponds to the low-$T$ anomaly in $M(T)$ and should be caused by a change of magnetic state of the Er$^{3+}$ moments. With applying a $c$-axis field, this peak becomes wider with increasing field for $\mu_0H <$ 0.1 T; for field above 1 T, the peak is strongly enhanced and behaves like the feature of a first-order phase transition, and moves to higher temperature with further increasing field. In addition, it is found that the transition temperatures agree well with the magnetic transition at $T_2$ in the $M(T)$ curves. In the case of $H \parallel a$, the specific-heat peak is gradually suppressed and nearly smeared out at 1 T. Besides, the broad hump becomes stronger and moves to higher temperature with increasing field, behaving like a Schottky anomaly.

\subsection{Thermal conductivity}

\begin{figure}
\includegraphics[clip,width=6.5cm]{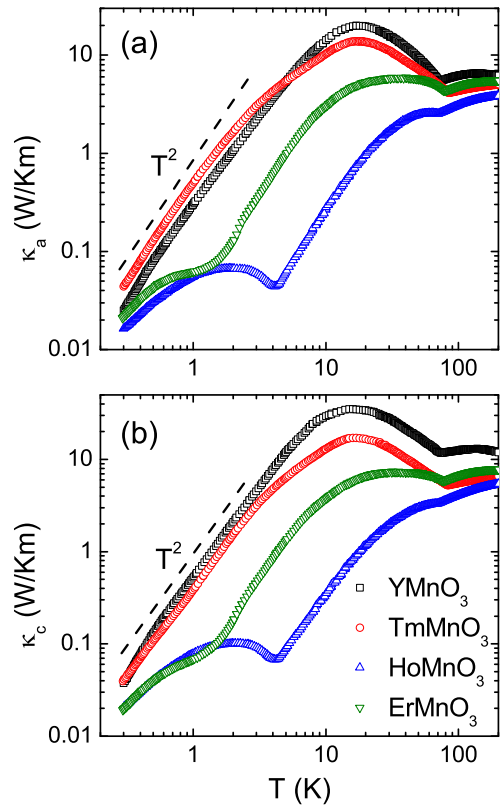}
\caption{(color online) Temperature dependence of the $a$-axis and the $c$-axis thermal conductivities of ErMnO$_3$ single crystals, compared with the results of YMnO$_3$, TmMnO$_3$ and HoMnO$_3$.\cite{Wang10, Wang12} The dashed lines show a $T^2$ temperature dependence.}
\end{figure}

Figure 9 shows the temperature dependence of the $a$-axis and the $c$-axis thermal conductivities ($\kappa_a$ and $\kappa_c$) of ErMnO$_3$ single crystals in zero field, in comparison with those of YMnO$_3$, TmMnO$_3$ and HoMnO$_3$.\cite{Wang10, Wang12} It is found that the $\kappa(T)$ of all these materials show a clear dip (or kink) at N\'eel temperature of the Mn$^{3+}$ moments, caused by a drastic phonon scattering of the critical spin fluctuations.\cite{Zhao_CVO, Song_CFO} At low temperatures, the $\kappa(T)$ behave rather differently with the variation of $R^{3+}$ ions. For YMnO$_3$ and TmMnO$_3$, the $\kappa(T)$ show clear phonon peaks at temperatures of 10--20 K. While in the case of HoMnO$_3$ and ErMnO$_3$, the phonon peaks are wiped out and the $\kappa$ show much smaller magnitudes at lower temperatures. In detail, a strong dip appears at $\sim$ 4 K in the $\kappa(T)$ of HoMnO$_3$, which is related to the AF ordering of Ho$^{3+}$ moments.\cite{Nature04, Wang10} While in the case of ErMnO$_3$, the $\kappa(T)$ shows a distinct curvature at 1--2 K, which should be related to the magnetic-order transition of Er$^{3+}$ moments. Magnetic excitations, either magnons or short-range fluctuations, can act as scatterers of phonon. In YMnO$_3$, the Y$^{3+}$ ions are nonmagnetic. While for TmMnO$_3$, the Tm$^{3+}$ moments do not form long-range order till 0.3 K.\cite{Yen07, Lorenz13, Wang12} Because of lacking magnon excitations from the $R^{3+}$ spin system in YMnO$_3$ and TmMnO$_3$, it is understandable that the low-$T$ $\kappa$ of YMnO$_3$ and TmMnO$_3$ are significantly larger than those of HoMnO$_3$ and ErMnO$_3$. That is, the magnetic excitations of rare-earth sublattice in HoMnO$_3$ and ErMnO$_3$ can strongly scatter phonons at low temperatures.

In passing, it is notable that the $\kappa(T)$ of YMnO$_3$ and TmMnO$_3$ at subKelvin temperatures approximate $T^2$ rather than $T^3$ behavior, suggesting the microscopic phonon scattering is not negligible also for these two materials, although the scattering is much weaker than that in HoMnO$_3$ and ErMnO$_3$. Previous inelastic neutron scattering experiment on YMnO$_3$ has revealed strong spin fluctuations at temperatures as low as 5 K because of the geometrical frustration of Mn$^{3+}$ moments.\cite{J-Park03} Probably this kind of spin fluctuations existing at low temperatures can scatter phonons. In this regard, the theoretical analysis for describing this scattering effect is called for.

\begin{figure}
\includegraphics[clip,width=8.5cm]{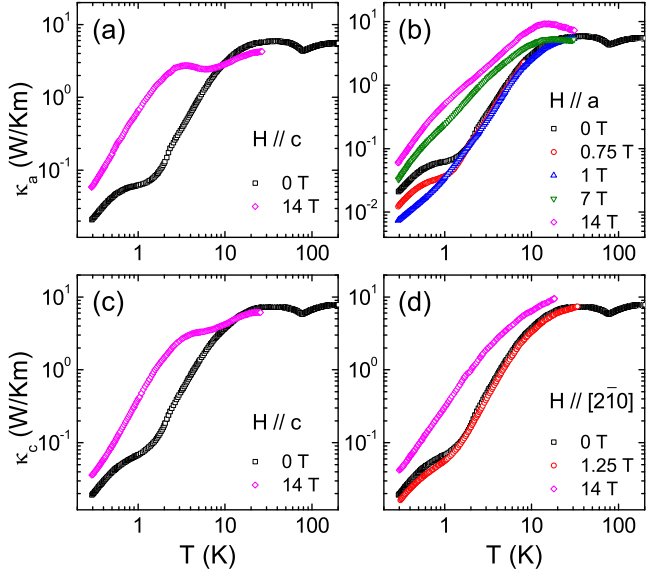}
\caption{(color online) Temperature dependence of the $a$-axis and the $c$-axis thermal conductivities of ErMnO$_3$ single crystals in magnetic fields.}
\end{figure}

Figure 10 shows the $\kappa(T)$ of ErMnO$_3$ single crystals in magnetic fields parallel to or perpendicular to the $c$ axis. With applying a 14 T field along the $c$ axis, as shown in Figs. 10(a) and 10(c), the low-$T$ $\kappa$ are strongly enhanced because of the suppression of magnetic excitations in high field; Besides, both the $\kappa_a(T)$ and $\kappa_c(T)$ exhibit a clear shoulder-like feature at the temperature range of 5 to 10 K. A similar phenomenon has been observed in the case of HoMnO$_3$ and TmMnO$_3$.\cite{Wang10, Wang12} For $H \parallel a$ or [2$\bar{1}$0], as shown in Figs. 10(b) and 10(d), a low magnetic field can clearly suppress the $\kappa$ at subKelvin temperatures. The curvature in $\kappa(T)$ extends to lower temperatures with increasing field, which has some correspondence with the results of specific heat for $H \parallel a$. With increasing field to 14 T, the low-$T$ $\kappa$ get a significant recovery; in particular, a clear phonon peak appears at $\sim$ 14 K in the $\kappa_a(T)$ for 14 T $\parallel a$.

\begin{figure}
\includegraphics[clip,width=8.5cm]{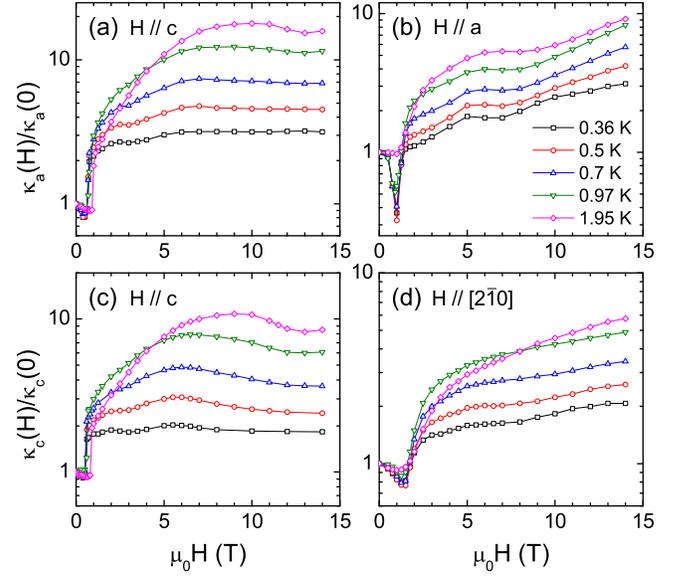}
\caption{(color online) Magnetic-field dependence of thermal conductivity of ErMnO$_3$ single crystals at $T <$ 2 K.}
\end{figure}

Figure 11 shows detailed magnetic-field dependence of thermal conductivities at $T <$ 2 K. For $H \parallel c$, the $\kappa(H)$ display a slight decrease at low field and a step-like increase at 0.5--0.8 T, as shown in Figs. 11(a) and 11(c). The $\kappa$ increases gradually with further increasing field, and then exhibits a broad-valley-like feature around 12 T. It is found that these changes of $\kappa(H)$ for $H \parallel c$ have good correspondence with the anomalies in the $c$-axis $M(H)$ at $H_{c1}$ and $H_{c2}$. In the case of $H \parallel ab$, as shown in Figs. 11(b) and 11(d), a dip-like feature is observed in the $\kappa(H)$ isotherms at $\sim$ 1 and 1.25 T for $H \parallel a$ and [2$\bar{1}$0], respectively, and the dip fields $H_c$ are nearly independent of temperature. A similar phenomenon, associated with a spin re-orientation of Mn$^{3+}$ sublattice with $90^{\circ}$ rotations in the $ab$ plane, has also been detected in the case of HoMnO$_3$ and TmMnO$_3$.\cite{Wang10, Wang12, Nature04} Further increasing field can effectively suppress the magnetic excitations and the spin-phonon scattering is weakened, leading to the gradual recovery of $\kappa$.

\subsection{Ground state and magnetic transitions}

\begin{figure}
\includegraphics[clip,width=8.5cm]{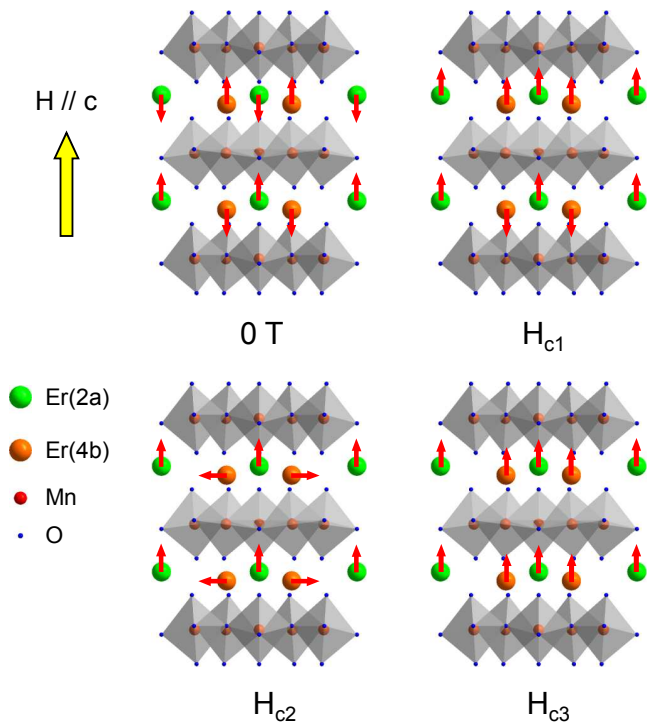}
\caption{(color online) Magnetic structures of rare-earth moments in ErMnO$_3$ at low temperatures and those in a $c$-axis field.}
\end{figure}

Based on above data, we proposed that the ground-state magnetic structures of Er$^{3+}$ moments in ErMnO$_3$ should not be the ferrimagnetic $P6_3c'm'$, which can undergo only one field-induced transition in $H \parallel c$. Among the four one-dimensional states, only $P6'_3cm'$ can be used to explain the experimental results. As shown in Fig. 12, with applying a $c$-axis field, the Er$^{3+}$(2a) moments are polarized along the $c$ axis at $H_{c1}$, and the magnetic excitations can be effectively suppressed, leading to quick increase of $\kappa$. Then, the Er$^{3+}$(4b) moments can undergo a spin-flop transition at $H_{c2}$ before coming into the spin polarized state at $H_{c3}$. Note that the transition at $H_{c2}$ does affect the $\kappa$, as shown in Figs. 9(a) and 9(c). Finally, the Er$^{3+}$(4b) moments are fully polarized at $H_{c3}$.

In addition, it was reported that the magnetic transition of Er$^{3+}$(4b) can be triggered by a change of Mn$^{3+}$ order and vice versa because of a rather strong $3d-4f$ interaction between Mn$^{3+}$ and Er$^{3+}$(4b) sublattices, while there is no coupling between the Mn$^{3+}$ and Er$^{3+}$(2a) sublattices.\cite{Meier12, Chaix14} Therefore, it is deduced that the magnetic structure of Mn$^{3+}$ should have corresponding spin reorientation at $H_{c2}$ and $H_{c3}$, which needs a further study using more specialized methods like Second Harmonic Generation (SGH).

In the case of $H \parallel ab$, it is suggested that all the Er$^{3+}$ moments should be polarized at a low transition field with a synchronous spin reorientation of Mn$^{3+}$ moments, resulting in a dip-like decline of $\kappa$ at the magnetic transitions.

\subsection{A comparison of the rare-earth ordering in HoMnO$_3$ and ErMnO$_3$}

In the family of $h$-$R$MnO$_3$ with magnetic rare-earth elements, there are a wealth of magnetic orders, magnetic phase transitions and magnetoelectric physics, and different $R$ can lead to rather different behaviors in these physical properties. Among these members, TmMnO$_3$ has been understood pretty well. In TmMnO$_3$, the Mn$^{3+}$ moments form the long-range order with $P6'_3c'm$ spin configuration at $T_N$ = 84 K. At low temperatures, there is no evidence of the long-range order of Tm$^{3+}$ moments till 0.3 K, but it was found that the Tm$^{3+}$(4b) moments form short-range AF order below $T_N$.\cite{Yen07, Skumryev09, Salama09, Wang12, Lorenz13} With applying a $c$-axis magnetic field in the ground state, the Tm$^{3+}$(4b) moments can be polarized along the magnetic field together with a spin re-orientation of the Mn$^{3+}$ sublattice from $P6'_3c'm$ to $P6_3c'm'$ state.\cite{Yen07, Wang12, Lorenz13}

ErMnO$_3$ and HoMnO$_3$ were found to exhibit very complicated $H-T$ phase diagrams.\cite{Yen05, Meier12} As mentioned above, it was originally proposed that at the ground state of ErMnO$_3$, the Er$^{3+}$(4b) and Er$^{3+}$(2a) sublattices have spin configuration of ferrimagnetic $P6_3c'm'$.\cite{Meier12} However, the magnetization and thermal conductivity data in present work indicate that the ground state of Er$^{3+}$(4b) and Er$^{3+}$(2a) sublattices are more likely $P6'_3cm'$. The $a$-axis field can induce one transition of Er$^{3+}$ sublattice, accompanied with the spin re-orientation of Mn$^{3+}$ sublattice, while the $c$-axis field can induce three magnetic transitions.

\begin{figure}
\includegraphics[clip,width=8.5cm]{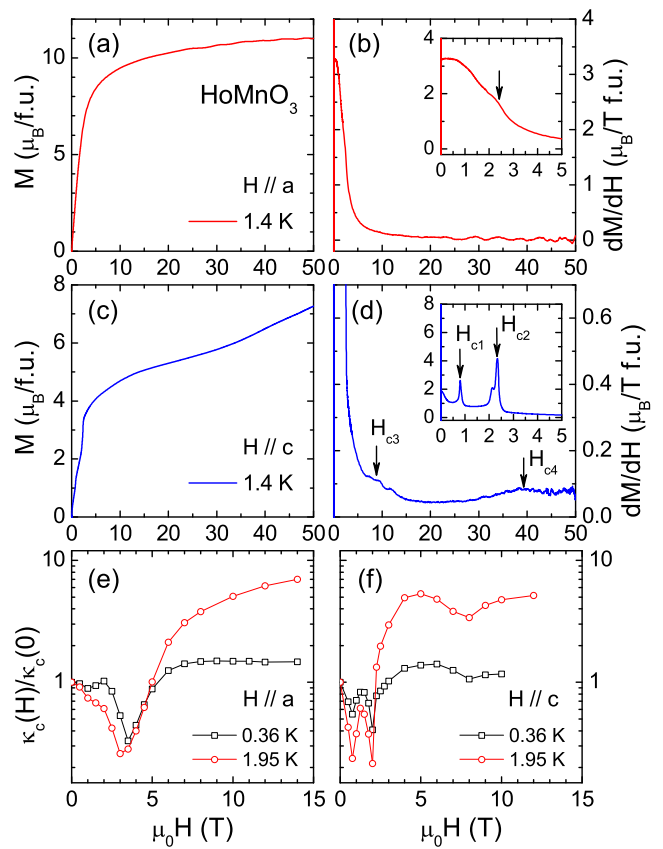}
\caption{(color online) (a,c) Low-temperature magnetization curves $M(H)$ of HoMnO$_3$ obtained in pulsed high magnetic field along the $a$ or $c$ axis. (b,d) The differential results of $dM/dH$ for the $a$- and $c$-axis $M(H)$. Inset to panels (b,d): low-field data of $dM/dH$. The arrows in panels (b,d) indicate the magnetic transitions. (e,f) Low-temperature field dependence of $\kappa_c$ for field along the $a$ or $c$ axis. Data are taken from Ref. \onlinecite{Wang10}.}
\end{figure}

The phase diagram of HoMnO$_3$ has been studied extensively and was found to be rather controversial. The magnetic order of Mn$^{3+}$ moments forms at $T_N$ = 76 K. At lower temperatures, the Mn$^{3+}$ sublattices can undergo two spin re-orientations at $T_{SR}$ = 33 K and $T_{Ho}$ = 5.4 K, and the magnetic structure of Mn$^{3+}$ in the ground state was confirmed to be $P6_3cm$.\cite{Fiebig2000, Sugie02, Fiebig03, Vajk05, Nandi08} However, the magnetic orders of Ho$^{3+}$ moments are still under debate. One supposition is: both Ho$^{3+}$(2a) and Ho$^{3+}$(4b) moments form AF $P6'_3cm'$ state at the temperature regime of $T_{Ho}$ $\leq T <$ $T_{SR}$; at lower temperatures, the Ho$^{3+}$ moments undergo a magnetic-structure transition from $P6'_3cm'$ to $P6_3cm$ state at $T_{Ho}$; hence, the Ho$^{3+}$(4b) moments are ordered along the $c$ axis with AF intra- and inter-plane couplings and the Ho$^{3+}$(2a) moments are disordered in the ground state.\cite{Vajk05, Nandi08, Fabreges09, Nature04} Another picture is: the magnetic order of Ho$^{3+}$ moments does not change at $T < T_{Ho}$, and both Ho$^{3+}$(2a) and Ho$^{3+}$(4b) sublattices have the long-range magnetic order with $P6'_3cm'$ state in the ground state.\cite{Brown06} In addition, the magnetic phase diagram of HoMnO$_3$ sketched by dielectric constant measurement indicates that the ground state of HoMnO$_3$ can undergo three field-induced magnetic transitions in a $c$-axis field up to 8 T.\cite{Yen05} Our previous thermal conductivity measurement indicated three dips in the low-$T$ $\kappa(H)$ curves with $H \parallel c$,\cite{Wang10} as shown in Fig. 13(f). Two low-field sharp dips can be clearly associated with magnetic transitions, but it was not confirmed in Ref. \onlinecite{Wang10} whether the broader one at 8 T is also associated with a magnetic transition because of lacking the high-field magnetization measurement.

\begin{figure}
\includegraphics[clip,width=8.5cm]{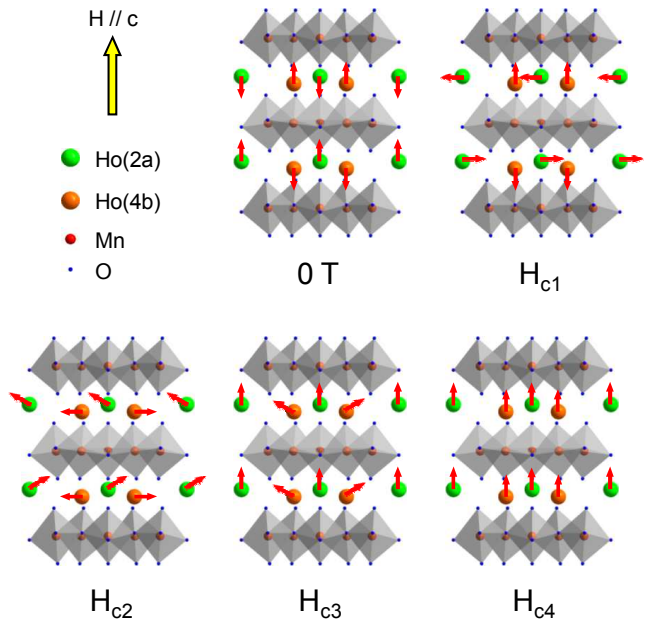}
\caption{(color online) Magnetic structures of rare-earth moments in HoMnO$_3$ at low temperatures and those in a $c$-axis field.}
\end{figure}

In the present work, the high-field magnetization and thermal conductivity of ErMnO$_3$ are able to reveal some new information on the magnetic ground state and field-induced transitions. In this regard, it is really useful to carry out the high-field experiments for other $h$-$R$MnO$_3$. Here, we also measured the magnetization of HoMnO$_3$ in high fields up to 50 T, as shown in Figs. 13(a--d). The data indicate that the magnetic structure actually undergoes four transitions in the $c$-axis magnetic field. Three low-field transitions are coincided with those in the $\kappa(H)$ and dielectric constant measurements and the fourth transition is located at very high field. With applying an $a$-axis magnetic field, both the $M(H)$ and $\kappa(H)$ curves show an anomaly at the same transition field. Based on this, it is suggested that the magnetic structure of Ho$^{3+}$ moments in HoMnO$_3$ should adopt $P6'_3cm'$ state, which is similar to the case of ErMnO$_3$. But the magnetization process in $H \parallel c$ for these two materials actually have some difference. Unlike the case of ErMnO$_3$, Ho$^{3+}$(2a) in HoMnO$_3$ should have a two-step magnetization in $H \parallel c$. As shown in Fig. 14, the magnetic transitions of Ho$^{3+}$ moments in $H \parallel c$ can be understood as following: at $H_{c1}$, the Ho$^{3+}$(2a) moments first undergo a 90$^\circ$ spin-flop transition from the $c$ axis to the $ab$ plane; with increasing field to $H_{c2}$, the Ho$^{3+}$(4b) moments undergo a 90$^\circ$ spin-flop transition from the $c$ axis to the $ab$ plane; then, the Ho$^{3+}$(2a) moments can be fully polarized along the $c$ axis at $H_{c3}$; finally, all Ho$^{3+}$ sublattices are fully polarized along the field at $H_{c4}$. With applying magnetic field in the $ab$ plane, the Ho$^{3+}$ moments can be polarized at transition field $H_{c}$, simultaneously, the Mn$^{3+}$ sublattice can undergo a spin re-orientation. With the high-field magnetization data, it is clear that the 8 T minimum in the low-$T$ $\kappa(H)$ curves ($H \parallel c$) indeed indicates a field-induced magnetic transition. In addition, we can make step forward in understanding the nature of ground state and the magnetic transitions of HoMnO$_3$. In addition, these results indicate that the spin anisotropy of Ho$^{3+}$ may not be Ising like.

These differences in the ground states and magnetic transitions among $h$-$R$MnO$_3$ are directly caused by different rare-earth ions, which mainly results in different magnetic interactions between the Mn$^{3+}$ and $R^{3+}$(4b) or $R^{3+}$(2a) moments. For example, the differences in the magnetization process of Ho$^{3+}$(2a) and Er$^{3+}$(2a) should be attributed to a difference in anisotropy energy. That is, the $R^{3+}$(2a)-Mn$^{3+}$ interaction in HoMnO$_3$ is more stronger than that in ErMnO$_3$, which is directly caused by different rare-earth ions. In addition, the $\kappa(H)$ of HoMnO$_3$ with $H \parallel c$ show clear different features at the transition fields of Ho$^{3+}$(2a), in comparison with the case of ErMnO$_3$. In detail, the $\kappa(H)$ of HoMnO$_3$ in $H \parallel c$ show two dip-like feature at $H_{c1}$ and $H_{c2}$, which indicates that a synchronous spin re-orientation of Mn$^{3+}$ should be triggered by the field-induced transition of Ho$^{3+}$(2a) moments. In contrast, the $\kappa(H)$ of ErMnO$_3$ with $H \parallel c$ show rather weak field dependence at low fields followed by a step-like increase at the polarization transition of the Er$^{3+}$(2a) moments, which indicates there is no accompanied spin re-orientation of Mn$^{3+}$ with the magnetic transition of Er$^{3+}$(2a). These difference reveals a difference in the R$^{3+}$(2a)-Mn$^{3+}$ interaction for HoMnO$_3$ and ErMnO$_3$.

\section{SUMMARY}

In conclusion, we carried out detail measurements on magnetization, specific heat and thermal conductivity of ErMnO$_3$ at low temperatures and in high magnetic fields. It is found that the magnetic structure of ErMnO$_3$ undergoes three transitions at 0.8, 12 and 28 T for $H \parallel c$. This indicates that the ferrimagnetic $P6_3c'm'$ spin configuration cannot be the actual ground state of Er$^{3+}$ sublattices. Meanwhile, $M(H)$, $C_p(T)$ and $\kappa(H)$ for $H \parallel a$ all show clear changes around 1 T, which confirms the presence of a magnetic transition. Based on present experiments, we conclude that the ground state of Er$^{3+}$ moments is likely $P6'_3cm'$. In addition, the difference in $M(H)$ and $\kappa(H)$ of HoMnO$_3$ and ErMnO$_3$ can be well understood by a difference in the R$^{3+}$(2a)-Mn$^{3+}$ interaction caused by different rare-earth ions. It is concluded that varying rare-earth ions in $h$-$R$MnO$_3$ is sensitive in control of the magnetic interactions between Mn$^{3+}$ and $R^{3+}$(4b) or $R^{3+}$(2a) moments, which offers an effective way to improve the physical performance of these materials.

\begin{acknowledgements}

This work was supported by the National Natural Science Foundation of China (Grant Nos. 11374277, U1532147, 11574286, 11404316), the National Basic Research Program of China (Grant Nos. 2015CB921201, 2016YFA0300103), the Opening Project of Wuhan National High Magnetic Field Center (Grant No. 2015KF21), and the Innovative Program of Development Foundation of Hefei Center for Physical Science and Technology.

\end{acknowledgements}

\end{document}